\documentclass[twocolumn,showpacs,preprintnumbers,amsmath,amssymb]{revtex4}

\usepackage{graphicx}
\usepackage{dcolumn}
\usepackage{bm}

\begin{document}

\title{Magnetic anisotropy switching in (Ga,Mn)As with increasing hole concentration}

\author{K. Hamaya,* T. Watanabe, T. Taniyama,$^{a)}$ A. Oiwa,$^{b),}$** Y. Kitamoto, and Y. Yamazaki}
\affiliation{Department of Innovative and Engineered Materials, Tokyo Institute of
Technology,\\ 4259 Nagatsuta, Midori-ku, Yokohama 226-8502, Japan.}%
\affiliation{$^{a)}$Materials and Structures Laboratory, Tokyo Institute of Technology,\\ 4259 Nagatsuta, Midori-ku, Yokohama 226-8503, Japan.}
\affiliation{$^{b)}$PRESTO, Japan Science and Technology Agency, 4-1-8 Honcho, Kawaguchi 332-0012, Japan.}


\date{\today}
\begin{abstract}
{We study a possible mechanism of the switching of the magnetic easy axis as a function of hole concentration in (Ga,Mn)As epilayers. In-plane uniaxial magnetic anisotropy along [110] is found to exceed intrinsic cubic magnetocrystalline anisotropy above a hole concentration of $p =$ 1.5 $\times$ 10$^{21}$ cm$^{-3}$ at 4 K. This anisotropy switching can be realized by post-growth annealing, and the temperature-dependent ac susceptibility is significantly changed with increasing annealing time. On the basis of our recent scenario [Phys. Rev. Lett. {\bf 94}, 147203 (2005); Phys. Rev. B {\bf 73}, 155204 (2006)], we deduce that the growth of highly hole-concentrated cluster regions with [110] uniaxial anisotropy is likely the predominant cause of the enhancement in [110] uniaxial anisotropy at the high hole concentration regime. We can clearly rule out anisotropic lattice strain as a possible origin of the switching of the magnetic anisotropy.

\vspace{5mm}
\scriptsize{ *Present address: Institute of Industrial Science, The University of Tokyo, 4-6-1 Komaba, Meguro-ku, Tokyo 153-8505, Japan. E-mail:  hamaya@iis.u-tokyo.ac.jp

**Present address: Department of Applied Physics, The University of Tokyo, 7-3-1 Hongo, Bunkyo-ku, Tokyo 113-8656, Japan.}}

\end{abstract}

\maketitle

\section{INTRODUCTION}

Ferromagnetism in III-V magnetic alloy semiconductors has been studied in many theoretical and experimental approaches,\cite{Ohno2,Matsukura,Dietl2,Abolfath,Keavney} invoked to $p-d$ exchange interaction between hole carriers and magnetic impurities. One of the most fascinating magnetic features in (Ga,Mn)As epilayers is the temperature variation of the magnetic anisotropy from $\left\langle 100 \right\rangle$ to [110] with increasing temperature,\cite{Welp,Liu,Hamaya2,Sawicki} where the origin of [110] uniaxial anisotropy remains still unclear. Also, the mechanism of the anisotropy switching from $\left\langle 100 \right\rangle$ to [110] with increasing temperature had not been elucidated, although theoretical works\cite{Dietl2,Abolfath} have made an attempt to correlate the magnetic anisotropy with the shape of the Fermi surface of valence hole subbands.

In regard to this point, we have recently proposed another two-phase scenario to explain the magnetic properties, where two ferromagnetic phases of [110] uniaxial anisotropic clusters and $\left\langle 100 \right\rangle$ matrix coexist in a (Ga,Mn)As epilayer:\cite{HamayaPRL,HamayaPRB} [110] uniaxial anisotropic clusters have relatively high hole concentration in the epilayer. Using this model, we can clearly explain the origin of the temperature-dependent anisotropy switching: with increasing temperature, the ferromagnetic $\left\langle 100 \right\rangle$ matrix phase becomes paramagnetic, and, the [110] uniaxial anisotropic clusters dominate the magnetism of (Ga,Mn)As epilayers near the Curie temperature.\cite{HamayaPRL,HamayaPRB} This cluster/matrix scenario is based on the fact that the low-temperature grown (Ga,Mn)As epilayers have point defects such as Mn interstitials (Mn$_\mathrm{I}$), \cite{Yu,Blinowski} and the Mn$_\mathrm{I}$ donors are diffused by post-growth low-temperature annealing.\cite{Ku,Edmonds} Also, the presence of energetically favorable Mn clusters, e.g., Mn$_\mathrm{Ga}$-Mn$_\mathrm{Ga}$ and Mn$_\mathrm{Ga}$-Mn$_\mathrm{I}$-Mn$_\mathrm{Ga}$, is proposed by theoretical studies \cite{Mahadevan,Raebiger} and direct observation using scanning tunneling microscopy.\cite{Sullivan} Recently, Sato {\it et al.} theoretically suggested the presence of spinodal decomposition phases in magnetic semiconductors such as (Ga,Mn)As and (Ga,Mn)N, where the phases bring about a high Curie temperature.\cite{Sato} 

On the other hand, the effect of hole concentration ($p$) on the magnetic anisotropy has also been examined experimentally. For example, it has been reported that $\left\langle 100 \right\rangle$ magnetocrystalline anisotropy can be controlled by changing $p$ in micro-structured (Ga,Mn)As wires\cite{HamayaJJAP} and modulation-doped (Ga,Mn)As/(Ga,Al)As heterostructures,\cite{Liu2} indicating that $p$ is a predominant parameter in determining the cubic magnetocrystalline anisotropy. In terms of in-plane uniaxial anisotropy along [110], we have demonstrated that the [110] uniaxial anisotropy is enhanced in an annealed (Ga,Mn)As epilayer compared with the corresponding as-grown epilayer.\cite{Kato} Also, Sawicki {\it et al.}\cite{Sawicki2} suggested that the in-plane uniaxial anisotropy depends on $p$ using the $p-d$ Zener model on the assumption of a small trigonal distortion. Since such trigonal distortion is likely to be attributed to the distribution of Mn ions (Mn clusters), this assumption seems to be compatible with our scenario: the Mn clusters cause inhomogeneity of $p$ in an epilayer and the magnetic anisotropy. To obtain a common view about the $p$ dependence of the magnetic anisotropy, we hence need to reconsider the magnetic anisotropy problem as a function of $p$ on the basis of the cluster scenario.\cite{HamayaPRL,HamayaPRB}
\begin{figure}[t]
\includegraphics[width=8cm]{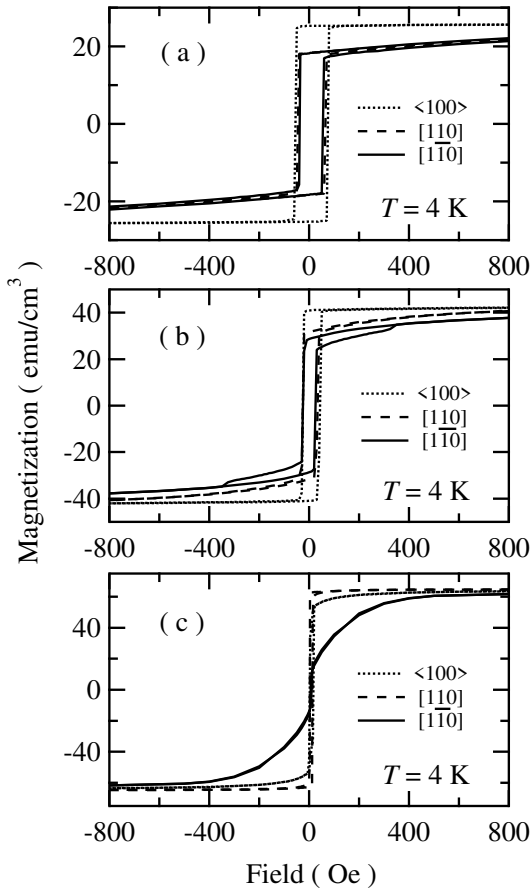}
\caption{$M-H$ curves of (Ga,Mn)As epilayers at 4 K. Magnetic field is applied parallel to either 
$\left\langle 100 \right\rangle$, [110], or [1$\overline{1}$0]. Hole concentrations ($p$) of three samples are 
(a) 3$\times$$10^{20}$ cm$^{-3}$, (b) 7$\times$$10^{20}$ cm$^{-3}$, and (c) 1.5$\times$$10^{21}$ cm$^{-3}$.}
\end{figure}

In this work, we show a clear manifestation of hole-concentration-dependent cubic and uniaxial anisotropy in (Ga,Mn)As epilayers, and find switching of the magnetic easy axis from $\left\langle 100 \right\rangle$ to [110] in the high hole concentration regime ($p\gtrsim$10$^{21}$ cm$^{-3}$) at 4 K. Comparison between ac susceptibility data of two separate epilayers with $p\sim$10$^{20}$ cm$^{-3}$ and $p\sim$10$^{21}$ cm$^{-3}$ reveals that [110] magnetic anisotropy arises from highly hole-concentrated (Ga,Mn)As clusters in the (Ga,Mn)As epilayer and such clusters are becoming predominant with increasing hole concentration.

\section{SAMPLES AND EXPERIMENT}
(Ga,Mn)As epilayers with a thickness of 100 or 200 nm were grown on a semi-insulating GaAs (001) substrate 
using low-temperature molecular beam epitaxy at 190 $-$ 240$^{\circ}$C. Prior to growing the (Ga,Mn)As, a 400-nm-thick GaAs buffer layer was grown in an As-stabilized condition at 590$^{\circ}$C. The Mn contents $x$ in the Ga$_{1-x}$Mn$_{x}$As epilayers were estimated to range from 0.024 to 0.088. After growth, some of the epilayers were annealed at 250$^{\circ}$C.\cite{Hayashi,Ku,Edmonds} Structural characterizations were performed by high resolution x-ray diffraction (HRXRD) using a Philips MRD. Magnetization was measured with a superconducting quantum interference device (SQUID) 
magnetometer. Ac susceptibility measurements were carried out using a physical property measurement system (PPMS).
 The hole carrier concentrations ($p$) which correspond to the ionized Mn acceptor concentration were measured with an electrochemical capacitance-voltage method at room temperature.\cite{Moriya} The samples we use show typical hole-mediated ferromagnetic properties; Curie temperature ($T_{c}$) and hole concentration ($p$) range from 55 K to 135 K and from 3$\times$$10^{20}$ cm$^{-3}$ to 1.5$\times$$10^{21}$ cm$^{-3}$, respectively, being  similar to those used in previous work reported by Ku {\it et al}.\cite{Ku}  

\section{RESULTS AND DISCUSSION}
Figure 1 shows the field-dependent magnetization ($M-H$ curve) of (Ga,Mn)As epilayers 
with different hole concentrations of (a) $p=$ 3$\times$$10^{20}$ cm$^{-3}$, (b) $p=$ 7$\times$$10^{20}$ cm$^{-3}$, and (c)  $p=$ 1.5$\times$$10^{21}$ cm$^{-3}$ at various field orientations at 4 K. 
For sample (a), no differences in the $M-H$ curves are seen for field directions along [110] and [1$\overline{1}$0] (Fig. 1(a)). In contrast, for samples (b) and (c), the differences between [110] and [1$\overline{1}$0] become clearer with increasing hole concentration (Figs. 1(b) and (c)), which is evidence of [110] uniaxial anisotropy. Also note that the remanent magnetization for [110] is larger than those for $\left\langle 100 \right\rangle$ in Fig. 1(c), indicating that the magnetic easy axis alters from [100] to a direction near [110] in the high hole concentration regime $p\gtrsim$$10^{21}$ cm$^{-3}$. 

To deduce the magnetic anisotropy constants, we assume that the $M-H$ curves along the hard axis can be modeled on coherent rotation.\cite{Welp} The magnetostatic energy of in-plane magnetized (Ga,Mn)As is generally given by $E =$ $K_{u}$sin$^{2}$($\varphi - $45$^\circ$) $+$ ($K_{1}$/4)sin$^{2}2\varphi$ $-$ $MH$cos($\varphi - \theta$), where {\it K$_{u}$} and {\it K$_{1}$} are the in-plane uniaxial and cubic anisotropy constants, respectively, {\it M} is the magnetization, {\it H} is the applied field strength, $\varphi$ is the magnetization direction, and $\theta$ is the applied field direction with respect to GaAs[100].\cite{Welp,Hrabovsky,Hamaya2} 
We fit a calculated $M-H$ curve in the field region from 0 Oe to 2 kOe using the conditions $\partial${\it E}/$\partial$$\varphi$ $=$ 0, $\partial$$^2${\it E}/$\partial$$\varphi$$^2$ $>$ 0, 
and $\theta =$ 135$^{\circ}$ ([1$\overline{1}$0]) in this formula.\cite{note} 
Figure 2 (a) plots {\it K$_{u}$} and {\it K$_{1}$} at 4 K as a function of $p$. 
{\it K$_{1}$} decreases from 3$\times$10$^{4}$ to 6$\times$10$^{3}$ (erg/cm$^{3}$) with increasing $p$. In contrast, {\it K$_{u}$} gradually increases from 1$\times$10$^{3}$ to 8$\times$10$^{3}$ (erg/cm$^{3}$) and becomes larger than {\it K$_{1}$} for a sample with $p =$ 1.5$\times$$10^{21}$ cm$^{-3}$, showing significant [110] uniaxial anisotropy. Our experimental data in Fig. 2 (a) clearly show a crossover of the values of {\it K$_{u}$} and {\it K$_{1}$} in the high hole concentration regime $p\gtrsim$$10^{21}$ cm$^{-3}$.
\begin{figure}[t]
\includegraphics[width=8cm]{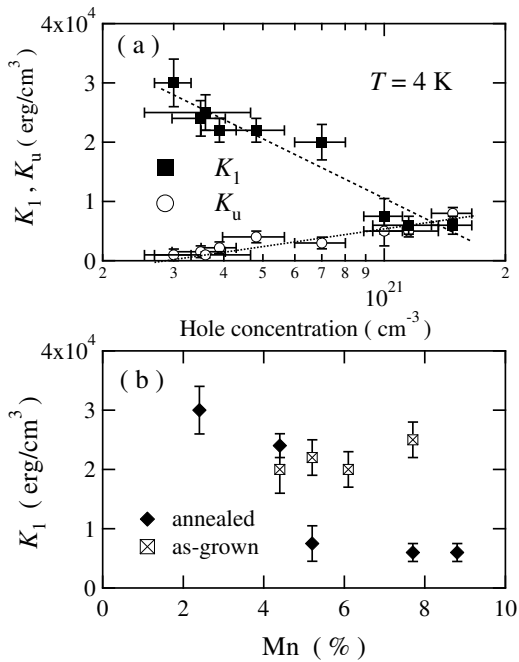}
\caption{(a) {\it K$_{u}$} and {\it K$_{1}$} estimated from $M-H$ curves at 4 K as a function of room-temperature hole concentration. (b) {\it K$_{1}$ } $vs$ Mn content at 4 K. Different symbols are used for as-grown and annealed samples. }
\end{figure}

Although $p$ seems to be the most appropriate parameter determining the magnetic anisotropy, another cause such as Mn content and lattice strain should be checked whether they could represent the experimental behavior. Here, Mn content is estimated from HRXRD measurements.\cite{Shen} As shown in Fig. 2 (b), it is clearly seen that the plots of {\it K$_{1}$} vs Mn content are rather scattered and show no systematic changes as a function of Mn content: even samples with the same Mn content show different {\it K$_{1}$} values depending on whether the sample is as-grown or annealed. Also, the magnetic anisotropy of (Ga,Mn)As was reported to be affected by the shape of valence hole subbands and lattice strain could be a possible cause of the change in the shape of the valence hole subbands.\cite{Dietl2,Abolfath} We have confirmed that all the epilayers used were grown coherently with tetragonal compressive strain and no misfit dislocation, and that no in-plane crystal asymmetry between [110] and [1$\overline{1}$0] is induced by increasing Mn contents or low temperature annealing, ensuring that the influence of the asymmetric lattice strain on the change in the values of {\it K$_{u}$} can be ignored. Welp {\it et al}. also showed that [110] uniaxial anisotropy field did not depend on epilayer thickness and the effects of (Ga,Mn)As surface and (Ga,Mn)As/GaAs(001) interface on [110] uniaxial anisotropy are not significant.\cite{Welp2} These experimental facts support our description that $p$ is most likely the predominant parameter in determining the magnetic anisotropy of (Ga,Mn)As rather than other parameters. 
\begin{figure}
\includegraphics[width=8.5cm]{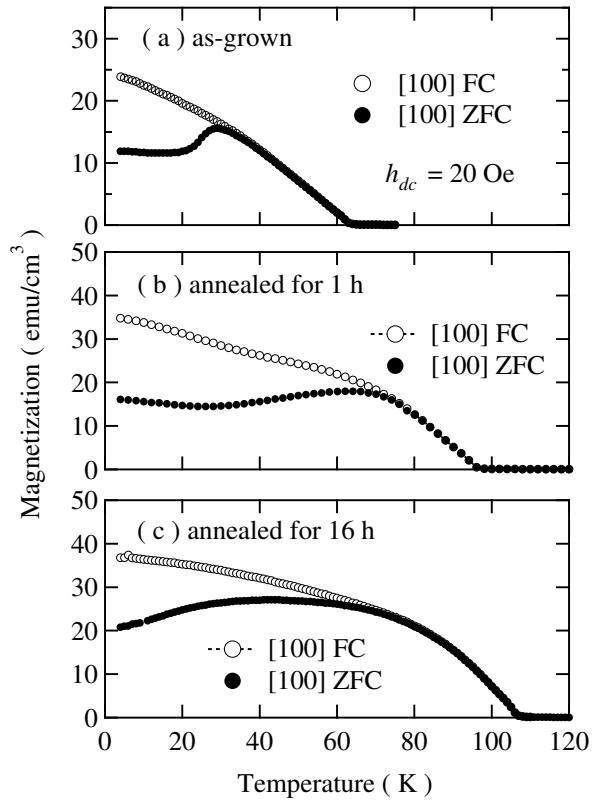}
\caption{Zero-field cooled and field cooled magnetizations for samples with various annealing time ((a) as-grown, (b) 1-hour annealed, (c) 16-hour annealed).}
\end{figure}

To gain additional insight into the origin of the change in the magnetic anisotropy with $p$, we examine the magnetic features of as-grown and low-temperature annealed samples (1-hour annealed and 16-hour annealed samples) with a constant Mn content. For the sample with a Mn content of 4.4\%, the [110] uniaxial anisotropy at 4 K and the $p$ were enhanced with increasing annealing time, being consistent with previous studies.\cite{Kato,Sawicki2} Also, no anisotropy switching with increasing temperature was observed for 16-hour annealed sample because of the marked [110] uniaxial anisotropy.\cite{HamayaPRB} For these samples, we measure the zero-field cooled (ZFC) and field cooled (FC) magnetizations. Figure 3 presents the ZFC and FC magnetizations of (a) as-grown sample, (b) 1-hour annealed sample, and (c) 16-hour annealed sample at $h_{dc}$ $=$ 20 Oe along [100]. Clear bifurcations of the ZFC and FC magnetizations are seen for all the samples. If there are no inhomogeneous magnetic phases and the magnetic structure consists of a single-domain as pointed out by Wang {\it et al.},\cite{Wang} we cannot expect the bifurcation of the ZFC and FC magnetizations for the samples.\cite{HamayaPRB} Thus, these bifurcations strongly support the presence of inhomogeneous magnetic phases in (Ga,Mn)As epilayers.\cite{Mahadevan,Raebiger,Sullivan,HamayaPRL,HamayaPRB,Sato} 

For our samples, since the single-domain model presented by Wang {\it et al.}\cite{Wang} is not realistic, we hereafter discuss the cluster/matrix model in our recent reports:\cite{HamayaPRL,HamayaPRB} (Ga,Mn)As epilayers that show two different $\left\langle 100 \right\rangle$ cubic anisotropy and [110] uniaxial anisotropy are composed of a matrix phase with relatively low hole concentration and highly-hole concentrated (Ga,Mn)As clusters embedded in the matrix. In this model, the distribution in $p$ is likely due to the inhomogeneous distribution of Mn ions.\cite{Mahadevan,Raebiger,Sullivan,HamayaPRL,HamayaPRB,Sato} In Fig. 4, we also show the temperature-dependent ac susceptibility ($\chi_{2}$"$- T$) curve for the same samples as in Fig. 3. An ac magnetic field of 6 Oe is small enough because features seen in the higher temperature regime are compatible with the magnetization data measured in Fig. 3: the high-temperature features in Fig. 4 are seen at around the temperature where the magnetization disappears in Fig. 3. As shown in our previous study,\cite{HamayaPRL} two peak-to-dip features at $\sim$38 K and $\sim$57 K in the $\chi_{2}$"$- T$ curve are seen for the as-grown sample, arising from the matrix and the clusters with different Curie temperatures.\cite{HamayaPRL,HamayaPRB} Hence, the feature seen in the lower temperature regime ($\sim$38 K) is associated with the Curie temperature of the matrix.\cite{HamayaPRL} For annealed samples, significant changes in the $\chi_{2}$" are found: the sample subjected to 1-hour annealing shows a shift of the low-temperature feature at 38 K towards further lower temperatures, while the high-temperature feature at 57 K shifts towards higher temperatures, causing multiple-dip feature in the $\chi_{2}$"$- T$ curve. After 16-hours annealing, the high-temperature feature in the $\chi_{2}$"$-T$ further changes and shifts towards higher temperature, and the low-temperature feature almost disappears. 
\begin{figure}
\includegraphics[width=8.5cm]{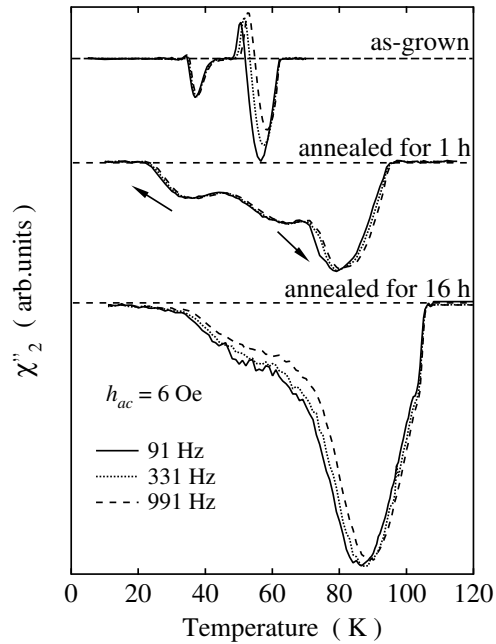}
\caption{Temperature dependent nonlinear ac susceptibility of the same samples used in dc magnetization measurements (Fig.3) at various ac frequencies.}
\end{figure}

We discuss the above systematic changes in the $\chi_{2}$" $- T$ curves, together with recent studies based on the presence of Mn clusters. Mahadevan {\it et al.}\cite{Mahadevan} theoretically suggested that Mn clusters, i.e., Mn$_\mathrm{Ga}$-Mn$_\mathrm{I}$-Mn$_\mathrm{Ga}$ complex, are formed favorably under As-rich low-temperature growth conditions. Edmonds {\it et al.}\cite{Edmonds} also proposed that the dissociation process of Mn$_\mathrm{I}$-Mn$_\mathrm{Ga}$ pairs is likely promoted in the region where Mn$_\mathrm{Ga}$ concentration is rather high and further low-temperature annealing diffuses the dissociated Mn$_\mathrm{I}$ ions. Since the number of Mn$_\mathrm{I}$ ions compensating holes decreases with annealing in such a region (highly hole-concentrated cluster region), the $p$ should increase with annealing, enhancing the Curie temperature as seen in the shift of the high-temperature feature in the $\chi_{2}$" $- T$ curve towards higher temperature. It is also likely that the volume of the highly hole-concentrated cluster region increases with annealing at the same time. On the other hand, some of the dissociated Mn ions which diffuse in the annealing process may be trapped again in the matrix phase and the dissociated Mn ions compensate holes, resulting in a decrease in the $p$ in the matrix. Therefore, the low-temperature feature in the $\chi_{2}$" $- T$ curve is suppressed with annealing and simultaneously the contribution of the cubic anisotropy which we attribute to the matrix phase is also reduced. The other dissociated Mn ions may segregate at the surface region\cite{Edmonds} and have no contribution to the change in the magnetic properties of the matrix phase. As a result, the increase in the $p$ in the cluster regions is more significant than the decrease in the $p$ in the matrix. These descriptions seem to be consistent with theoretical and experimental results on the Mn distribution in low-temperature grown (Ga,Mn)As epilayers.\cite{Yu,Blinowski,Mahadevan,Raebiger,Sullivan,Ku,Edmonds,HamayaPRL,HamayaPRB} 
In this context, the multiple-dip feature seen for 1-hour annealed sample is caused by an increase in the $p$ distribution, giving rise to the inhomogeneous magnetic properties. 

From these consideration, we conclude that highly hole-concentrated (Ga,Mn)As clusters formed during the growth are the origin of [110] uniaxial magnetic anisotropy on the basis of our recent picture of the cluster/matrix magnetic structure in (Ga,Mn)As. That is, the switching of the magnetic anisotropy as a function of $p$ originates from the change in the relative volume of highly hole-concentrated (Ga,Mn)As clusters.

\section{CONCLUSION}
In conclusion, we have observed a systematic change in {\it K$_{u}$} and {\it K$_{1}$} as a function of hole concentration ($p$), where the contribution of the [110] uniaxial anisotropy becomes larger than the cubic magnetocrystalline anisotropy at the high hole concentration regime ($p \sim$ 10$^{21}$ cm$^{-3}$) at 4 K. We have revealed that the contribution of [110] uniaxial anisotropy is becoming significant with increasing $p$ in the highly hole-concentrated (Ga,Mn)As clusters and the growth of the clusters. This mechanism is the most probable origin of the switching of magnetic easy axis with increasing $p$.

\vspace{2mm}
Authors thank Prof. H. Munekata of Tokyo Institute of Technology for the collaboration in the study of ferromagnetic semiconductors. 
The authors also thank Prof. H. Funakubo of Tokyo Institute of Technology for providing the opportunity to use their facility. 


\end{document}